\documentclass[12pt]{article}
\usepackage{times}
\usepackage{amsmath}
\usepackage{amsfonts}
\usepackage{amssymb}
\usepackage{graphicx}
\usepackage{cite}
\usepackage{color}
\usepackage{bm}
\usepackage[doublespacing]{setspace}

\newcommand{\upcite}[1]{\textsuperscript{\textsuperscript{\nocite{#1}\citenum{#1}}}}
\makeatletter
\renewcommand\@biblabel[1]{#1.}
\makeatother
\title{Frustrated superconductivity and intrinsic reduction of $T_c$ in trilayer nickelate} 

\author
{Qiong Qin,${}^{1,2,a}$, Jiangfan Wang,${}^{3,a}$ Yi-feng Yang${}^{1,2,4\ast}$\\
	\\
	\normalsize{${}^{1}$Beijing National Laboratory for Condensed Matter Physics and Institute of Physics,}\\
	\normalsize{Chinese Academy of Sciences, Beijing 100190, China}\\
	\normalsize{${}^{2}$University of Chinese Academy of Sciences, Beijing 100049, China}\\
	\normalsize{${}^{3}$School of Physics, Hangzhou Normal University, Hangzhou, Zhejiang 311121, China}\\
	\normalsize{${}^{4}$Songshan Lake Materials Laboratory, Dongguan, Guangdong 523808, China}\\
	\normalsize{${}^{a}$These authors contributed equally to this work.}\\
	\normalsize{$^\ast$Corresponding author. E-mail: yifeng@iphy.ac.cn}
}

\date{}

\begin{document} 

\baselineskip24pt

\maketitle

\begin{abstract}
Identifying the key factors controlling the magnitude of $T_c$ is of critical importance in the pursuit of high-temperature superconductivity. In cuprates, $T_c$ reaches its maximal value in trilayer structure, leading to the belief that interlayer coupling may help promote the pairing.  In contrast, for the recently discovered nickelate superconductors under high pressure, the maximum $T_c$ is reduced from about 80 K in the bilayer  La$_3$Ni$_2$O$_{7}$ to 30 K in the trilayer La$_4$Ni$_3$O$_{10}$. Motivated by this opposite trend, we propose an interlayer pairing scenario for the superconductivity of La$_4$Ni$_3$O$_{10}$. Our theory reveals intrinsic frustration in the spin-singlet pairing that the inner layer tends to form with both of the two outer layers respectively, leading to strong superconducting fluctuations between layers.  This explains the reduction of its maximum $T_c$ compared to that of the bilayer  La$_3$Ni$_2$O$_{7}$. Our findings support a fundamental distinction between multilayer nickelate and cuprate superconductors, and ascribe it to their different (interlayer versus intralayer) pairing mechanisms. Furthermore, our theory predicts extended $s^\pm$-wave gap structures in La$_4$Ni$_3$O$_{10}$, with varying signs and possible nodes on different Fermi pockets. We also find an intrinsic Josephson coupling with potentially interesting consequences that may be examined in future experiments. Our work reveals the possibility of rich novel physics in multilayer superconductors with interlayer pairing.
\end{abstract}

\section*{INTRODUCTION}

The recent discoveries of superconductivity in bilayer\upcite{Sun2023b,Hou2023,Zhang2023c} and trilayer\upcite{Sakakibara2023b,Li2024a,Zhu2023,Zhang2023m,Li2024} nickelates under high pressure have stimulated intensive experimental\upcite{Liu2023a,Yang2023f,Zhou2023a,Cui2023,Chen2023g,Xie2024,Chen2024a,Abadi2024,Dong2023,Dan2024,Wang2024PRX,Yuan2024,Kakoi2023b} and theoretical\upcite{Liu2023b,Liu2023f,Zhang2023b,Zhang2023d,Zhang2023e,Yang2023a,Yang2023b,Chen2023,Chen2023f,Luo2023,Christiansson2023,Shilenko2023,Wu2023a,Cao2023,Lechermann2023,Gu2023,Shen2023,Lu2023,Oh2023,Liao2023,Qu2023f,Jiang2023a,Qin2023b,Huang2023,Tian2023,Geisler2023,Lu2023c,Ryee2023,Chang2023,Quyang2023,Zheng2023,Heier2023,Talantsev2024,Botzel2024,Xue2024,Qu2023,Wang2024a,Luo2023f,Sakakibara2023,Wang2023g,Kaneko2024,Zhang2023Strong,Fan2023,Zhang2024a,Leonov2024,Tian2024,Wang2024b,LaBollita2024,Zhang2024b,Lu2024a,Yang2024a,Luo2024} debates concerning their pairing mechanisms as well as potential connections to the cuprate high-temperature superconductivity. While in-plane pairing has been well established in cuprates, both interlayer and intralayer pairing scenarios have been proposed to explain superconductivity in bilayer and trilayer nickelates. Latest resonant inelastic X-ray scattering (RIXS) measurements on La$_3$Ni$_2$O$_7$ at ambient pressure\upcite{Chen2024a} reported a dominant interlayer superexchange interaction $J$, which seems to support the interlayer scenario. However, straightforward experimental confirmation remains challenging due to the high pressure.

In this work, we point out that the different variation of the maximum $T_c$ in multilayer cuprate and nickelate superconductors might be an indicator of their possibly distinct pairing mechanisms. In cuprates, the maximum $T_c$ is the highest in trilayer systems. At ambient pressure, it rises from 97 K (single layer) to 135 K (trilayer) in Hg-based cuprates and from 30 K to 115 K in Bi-based cuprates\upcite{Scalapino2012a,Wang2023Science}. In contrast, the maximum $T_c$ is reduced from about 80 K in the bilayer nickelate La$_3$Ni$_2$O$_7$ to about 30 K in the trilayer nickelate La$_4$Ni$_3$O$_{10}$. To understand this difference, we propose an interlayer pairing scenario and perform numerical simulations on an effective low-energy model of the trilayer nickelate. Our calculations yield a maximum ratio $T_c/J\approx 0.02-0.03$, which is reduced by nearly a half compared to that of  $0.04-0.05$ in the bilayer model\upcite{Qin2023b}. This reduction is primarily attributed to strong superconducting fluctuations induced by some intrinsic frustration of the interlayer pairing fields,  which is absent in intralayer pairing, while imbalance between the inner and two outer layers may further suppress $T_c$ in the trilayer structure. An intrinsic Josephson coupling is also revealed that may have potentially interesting implications for future experimental investigations.

\section*{MATERIALS AND METHODS}

We start with the following low-energy effective two-orbital trilayer model based on experimental and first-principles considerations\upcite{Luo2024,Yang2023f,Li2017}, 
\begin{eqnarray}
	H&=&-\sum_{lijs}(t_{ij}+\mu\delta_{ij})c_{lis}^{\dagger}c_{ljs}-\sum_{lij}V_{ij}\left(c_{lis}^{\dagger}d_{ljs}+h.c\right) \label{eq:H} \nonumber\\
	&&-t_{\perp} \sum_{ais} \left(d_{0is}^\dagger d_{ais}+h.c.\right)+J\sum_{ai}\bm{S}_{0i}\cdot\bm{S}_{ai}, 
\end{eqnarray}
where $d_{lis}$ ($c_{lis}$) is the annihilation operator of $d_{z^2}$ ($d_{x^2-y^2}$) electrons with spin $s$ on site $i$ of the inner ($l=0$) or two outer ($l=\pm$) layers, $\bm{S}_{li}=\frac12\sum_{ss'}d_{lis}^{\dagger}\bm{\sigma}_{ss'}d_{lis'}$ is the $d_{z^2}$ spin density operator with $\boldsymbol{\sigma}$ being the Pauli matrices, the subscript $a=\pm$ denotes two outer layers, $t_{ij}$ and $\mu$ are the in-plane nearest-neighbor hopping and the chemical potential of $d_{x^2-y^2}$, $V_{ij}$ is the renormalized in-plane hybridization between nearest-neighbor $d_{z^2}$ and $d_{x^2-y^2}$ orbitals satisfying $V_{i,i+\hat{x}}=-V_{i,i+\hat{y}}=V$, $t_\perp$ is the renormalized interlayer hopping of $d_{z^2}$ quasiparticles, and $J$ is the $d_{z^2}$ interlayer superexchange interaction mediated by apical oxygens. Other small parameters are ignored for simplicity. The effect of the Hund's rule coupling is incorporated into the renormalized hybridization $V$ which is set as a free tuning parameter\upcite{Cao2023,Yang2023b}. The intralayer superexchange interaction is small as reported in latest RIXS and inelastic neutron scattering (INS) measurements\upcite{Chen2024a,Xie2024}, and thus ignored to focus on interlayer pairing only.

We employ the static auxiliary field Monte Carlo approach,  which has been widely applied to strongly correlated electron studies\upcite{Mayr2005a,Dubi2007,Karmakar2020,Pasrija2016,Dong2021a,Mukherjee2014,Liang2013,Dong2022PRB,Qin2023PRB,Han2010,Zhong2011,Singh2021}. Its application to the bilayer nickelate superconductor has successfully predicted the evolution of $T_c$ in good consistency with experiment\upcite{Qin2023b}. To simulate the trilayer model, we decouple the superexchange term\upcite{Coleman2015}:

\begin{equation}
	J\bm{S}_{0i}\cdot\bm{S}_{ai}\rightarrow \sqrt{2}\bar{\Delta}^a_i\Phi_{i}^a+\sqrt{2}\bar{\Phi}^a_{i}\Delta^a_i+\frac{8\bar{\Delta}^a_i\Delta^a_i}{3J},
\end{equation}
where $\Phi_{i}^a=\frac{1}{\sqrt{2}}\left(d_{0i\downarrow}d_{ai\uparrow}-d_{0i\uparrow}d_{ai\downarrow}\right)$ denotes the local interlayer $d_{z^2}$ spin singlet at site $i$ between the inner and $a$-th outer layers, and $\Delta_i^a$ is its corresponding pairing field. The static approximation ignores the imaginary time dependence of the auxiliary fields, $\Delta_i^a(\tau)\rightarrow \Delta_i^a=|\Delta_i^a|e^{i\theta_i^a}$, to avoid the severe sign problem, while the full thermodynamic fluctuations are retained beyond the uniform mean-field approximation. Hence, it is particularly suitable for studying the Berezinskii-Kosterlitz-Thouless (BKT) transition in two-dimensional superconductivity at finite temperatures\upcite{Berezinskii1972,Kosterlitz1973,Kosterlitz1974}.

We first integrate out all fermionic degrees of freedom to derive an effective action that solely depends on the pairing fields:
\begin{equation}
	\begin{split}
		S_{\rm eff}(\{\Delta^a_i\})=&\frac{8\beta}{3J}\sum_{ia}\bar{\Delta}_i^{a}\Delta_i^{a}-\sum_{i}\ln(1+e^{-\beta\tilde{\Lambda}_i})\\
		&-\sum_{i}\ln(1+e^{-\beta\Lambda_i}),
	\end{split}
\end{equation} 
where $\tilde{\Lambda}_i$ and $\Lambda_i$ are the eigenvalues of two matrices determined by the tight-binding parameters and the pairing configuration $\{\Delta^a_i\}$ (see supplemental information). Monte Carlo simulations are then performed to sample the probability distribution $p(\{\Delta_i^{a}\})=Z^{-1}e^{-S_{\rm eff}(\{\Delta_i^a\})}$, where $Z$ is the partition function serving as a normalization factor. Hereafter, we will set $t$ to unity as the energy unit and fix the chemical potential $\mu=-1.3$ to ensure that the total electron density varies only slightly around its nominal value 2.67. Due to computational limitations, all results presented are obtained on a $10\times 10$ trilayer lattice but the conclusions have been examined to be robust across other lattice sizes.

\section*{RESULTS}

\subsection*{Reduction of $T_c$}

\begin{figure}
	\begin{center}
		\includegraphics[width=14cm]{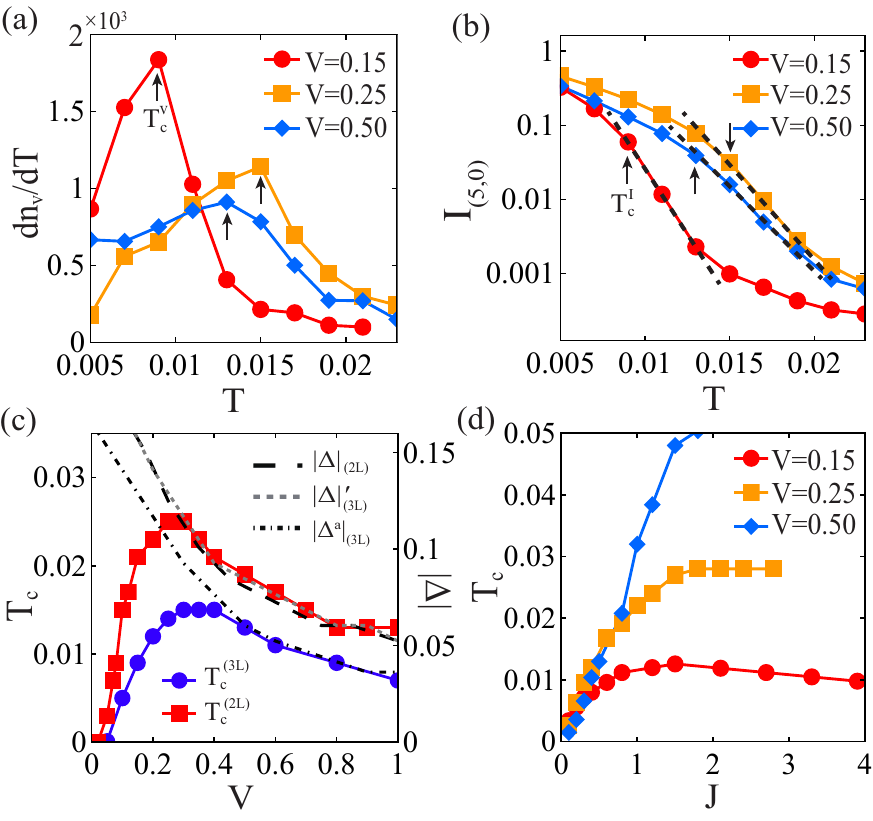}
	\end{center}
	\caption{\textbf{Superconducting transition temperature.} Temperature evolution of (A) the derivative $dn_{\rm v}/dT$ of the vortex number $n_{\rm v}$ and (B) the superconducting phase mutual information $I_{(5,0)}$ for three typical values of the hybridization parameter $V=0.15,0.25,0.50$ and $J=0.5$. The maximum of $dn_{\rm v}/dT$ reflects the BKT transition for two-dimensional superconductivity and defines the temperature scale $T_c^{\rm v}$, and the slope change (dashed lines) in $I_{(5,0)}$ at low temperatures marks the long-distance phase coherence and defines the temperature scale $T_c^{\rm I}$. They together define the superconducting transition temperature $T_{c}$. (C) Comparison of $T_c$ for trilayer and bilayer models as functions of the hybridization $V$ at $J=0.5$. Also shown are the most probable values of the pairing amplitudes $|\Delta^a|$ and $|\Delta|'=\sqrt{|\Delta^{+}|^2+|\Delta^{-}|^2}$ of  the trilayer model and $|\Delta|$ of the bilayer model. (D) Evolution of the trilayer $T_c$ as functions of the interlayer superexchange interaction $J$ for $V=0.15,0.25,0.50$.}
	\label{fig1}
\end{figure}

Theoretically, the BKT transition is characterized by a rapid increase of the vortex number $n_{\rm v}$\upcite{Drouin-Touchette2022}. Figure~\ref{fig1}A plots its temperature derivative $dn_{\rm v}/dT$ for three different values of $V$ at fixed $J=0.5$ and $t_\perp=0$. A peak is clearly seen for each curve that defines the temperature scale $T_c^{\rm v}$. To confirm that this indeed marks a superconducting transition, we further calculate the phase mutual information $I_{\mathbf{R}}$ between two local pairing fields of the largest distance $\mathbf{R}=(5,0)$ along the $x$ direction\upcite{Cover2006,Kraskov2004PRE,Khan2007,Belghazi2018PMLR,Poolel2019PMLR,Varanasi1999,Qin2023b}. As shown in Figure \ref{fig1}B ,  the phase mutual information displays several slope changes with lowering temperature in the semilog plot. Its rapid increase at intermediate temperature region marks the development of long-distance phase correlation, with a lower boundary $T_c^{\rm I}$ coinciding with $T_c^{\rm v}$ determined from the vortex number. We thus identify them as the superconducting transition temperature $T_c$ at which phase coherence is developed between local pairs on distant interlayer bonds.

Figure \ref{fig1}C compares the estimated $T_c$ for bilayer and trilayer models of the same parameters. Both vary nonmonotonically with the hybridization parameter $V$ and reach a maximum at around $V=0.3$. The nonmonotonic variation reflects the competition between in-plane phase coherence induced by the hybridization $V$ and local interlayer pairing determined mainly by the superexchange $J$. We find $T_c$ to be greatly reduced from bilayer to trilayer models. This observation is further supported by Figure \ref{fig1}D  where the trilayer $T_c$ is plotted against $J$ for three typical values of $V=0.15$, 0.25, 0.50. We obtain the largest slope in the small $J$ region, giving a maximum ratio $T_c/J\approx 0.02-0.03$ for realistic parameters. While for the bilayer model, our previous calculations yield a maximum $T_c/J\approx 0.04-0.05$\upcite{Qin2023b}. Interestingly, if we take heuristically the interlayer $J$ to be 150 meV from RIXS measurements (assuming the spin size $S\approx 1/2$) on La$_3$Ni$_2$O$_7$ at ambient pressure\upcite{Chen2024a}, these ratios immediately predict a maximum $T_c$ of about $70-90$ K for the bilayer system and $30-50$ K for the trilayer system, both consistent with experimental observations\upcite{Sun2023b,Zhang2023m}. In contrast, the intralayer $J$ is less than 10 meV at ambient pressure, whereas in cuprates it is much larger than the interlayer one\upcite{LeTacon2011}. 

In reality, the magnitude of the trilayer $T_c$ may be further reduced by other factors. First-principles calculations have revealed a subtle imbalance in La$_4$Ni$_3$O$_{10}$, with somewhat different hybridization $V$ and chemical potential $\mu$ between the inner and outer layers\upcite{Luo2024,Tian2024}. Our calculations show that both tend to suppress $T_c$ (see supplemental information). A small interlayer hopping may also influence $T_c$ but acts similarly on both models. We therefore conclude that the reduction of $T_c$ from bilayer to trilayer nickelates may be a genuine intrinsic property reflecting their interlayer pairing mechanism, different from the cuprates.

\subsection*{Frustrated superconductivity}

\begin{figure}
	\begin{center}
		\includegraphics[width=14cm]{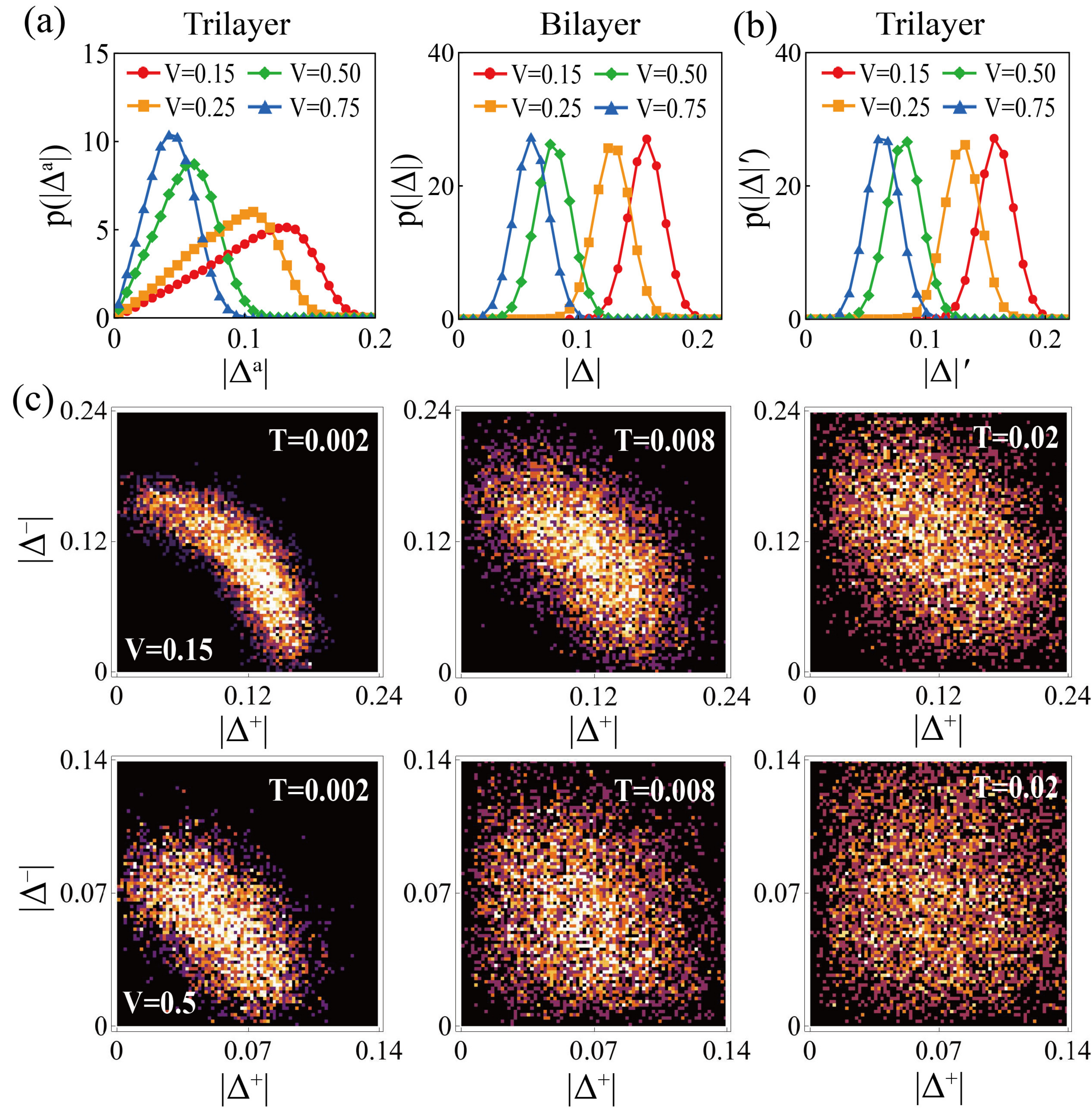}
	\end{center}
	\caption{\textbf{Interlayer superconducting fluctuations.} (A) Comparison of the probability distributions of the pairing amplitudes $p(|\Delta|^a)$ of the trilayer model (left) and $p(|\Delta|)$ of the bilayer model (right) for different values of $V=0.15,0.25,0.50,0.75$ at $T=0.002$. (B) Probability distribution of the total pairing amplitude $p(|\Delta|^{\prime})$ of the trilayer model, where $|\Delta|^{\prime}=\sqrt{|\Delta^{+}|^2+|\Delta^{-}|^2}$. (C) Intensity plot of the joint distribution of $|\Delta^{+}|$ and $|\Delta^{-}|$ of the trilayer model for $V=0.15,0.5$ and $T=0.002,0.008,0.02$, showing clear ring-shaped distribution at low temperatures and weak hybridizations. Other parameters are $t_\perp=0$ and $J=0.5$.}
	\label{fig2}
\end{figure}

To clarify the origin of the $T_c$ reduction, we present a comparative analysis of the probability distribution of the pairing field amplitudes in bilayer and trilayer models. As shown in Figure \ref{fig2}A , the trilayer $p(|\Delta^a|)$ exhibits a broad peak extending almost linearly to $|\Delta^a|=0$, in sharp contrast to that of the bilayer model. To understand this, we collect all Monte Carlo configurations of $|\Delta^a|$ of the trilayer model on an intensity plot in Figure \ref{fig2}C  for two typical values $V=0.15$ and 0.5. Quite unexpectedly, the data accumulate to form a ring shape at low temperatures. As the temperature increases, the ring gradually dissolves, resulting in nearly independent fluctuations of $|\Delta^+|$ and $|\Delta^-|$. Note that the suppression of the distribution near horizontal and vertical axes is due to the phase space limitation of the amplitude. For small $V=0.15$, the ring remains discernible even above $T_c$ due to preformed pairs in this parameter region\upcite{Keimer2015,Emery1995}. 

The ring shape suggests some internal symmetry between $|\Delta^\pm|$ and motivates us to define an overall pairing amplitude $|\Delta|'\equiv\sqrt{|\Delta^+|^2+|\Delta^-|^2}$. Interestingly, as shown in Figure \ref{fig2}B , the trilayer $p(|\Delta|')$ exhibits almost identical distribution as the bilayer $p(|\Delta|)$. To see this more clearly, we plot in Figure \ref{fig1}C  the most probable value of $|\Delta^a|$ and $|\Delta|'$ of the trilayer model and $|\Delta|$ of the bilayer model, and compare them with $T_c$ after a common scaling. Surprisingly, $|\Delta|'$ and $|\Delta|$ fall upon the same curve and track exactly the $T_c$ evolution of the bilayer model at large $V$, while $|\Delta^a|$ is systematically smaller and falls upon the $T_c$ curve of the trilayer model. Thus, the magnitudes of $T_c$ in this region are determined by the pairing amplitudes in both models, and its reduction in the trilayer model must originate from the fluctuation between $|\Delta^+|$ and $|\Delta^-|$. The superconductivity seems frustrated, where two outer layers compete to form the singlet with the inner layer to achieve an overall $|\Delta|'\approx |\Delta|$.

\subsection*{Internal symmetry and Josephson coupling}

To clarify the origin of the frustration, we expand the effective action up to fourth order by assuming uniform pairing fields for simplicity. This gives a Ginzburg-Landau (GL) free energy for the trilayer model:
\begin{equation}
	f_\text{GL}=c_1\Psi^\dagger \Psi+c_2(\Psi^\dagger \Psi)^2-2h \mathcal{T}^\Delta_{x},
\end{equation}
where $\Psi=(\Delta^{+},\Delta^{-})^T$ is a Higgs-like complex doublet order parameter, $\mathcal{T}^\Delta=\frac{1}{2}\Psi^\dagger \boldsymbol{\sigma}\Psi$ is its pseudospin, $c_{1/2}$ are two GL parameters determined by model details, and $h\propto t_\perp^2$ plays the role of an effective transverse field. Detailed derivations can be found in the supplemental information.

We first consider $t_\perp=0$ since it is strongly renormalized and proportional to the density of self-doped holes on $d_{z^2}$ ($\delta_h\sim 0.01$). This gives $h=0$ and we immediately see that the GL free energy has a global SU(2)$\times$U(1) symmetry associated with the  rotation of the pseudospin and an overall phase rotation. Both symmetries can be well understood from the original Hamiltonian (\ref{eq:H}), where the SU(2)  corresponds to the basis transformation of $(d_{+is},d_{-is})^T$ and $(c_{+is},c_{-is})^T$ between two outer layers, and the U(1) reflects their phase change relative to the inner layer (charge conservation). For $c_1<0$ and $c_2>0$, $\Psi^\dagger \Psi=|\Delta^{+}|^2+|\Delta^{-}|^2$ acquires a nonzero expectation $-c_1/2c_2$, indicating a finite mean-field value of the total amplitude $|\Delta|'$. However, the Mermin-Wagner theorem forbids the spontaneous continuous symmetry breaking at finite temperature, causing the ring-shaped joint distribution of $|\Delta^{+}|$ and $|\Delta^{-}|$ observed in Figure \ref{fig2}C. For the same reason, there is no phase coherence between two superconducting layers, since $\langle e^{-i(\theta^{+}-\theta^{-})}\rangle\propto\langle \bar{\Delta}^{+}\Delta^{-}\rangle =\langle \mathcal{T}^\Delta_x\rangle+i\langle \mathcal{T}^\Delta_y\rangle=0$, which is also observed in our simulations. The free energy has the same form for the bilayer $|\Delta|$, explaining their almost identical distributions.

\begin{figure}
	\begin{center}
		\includegraphics[width=14cm]{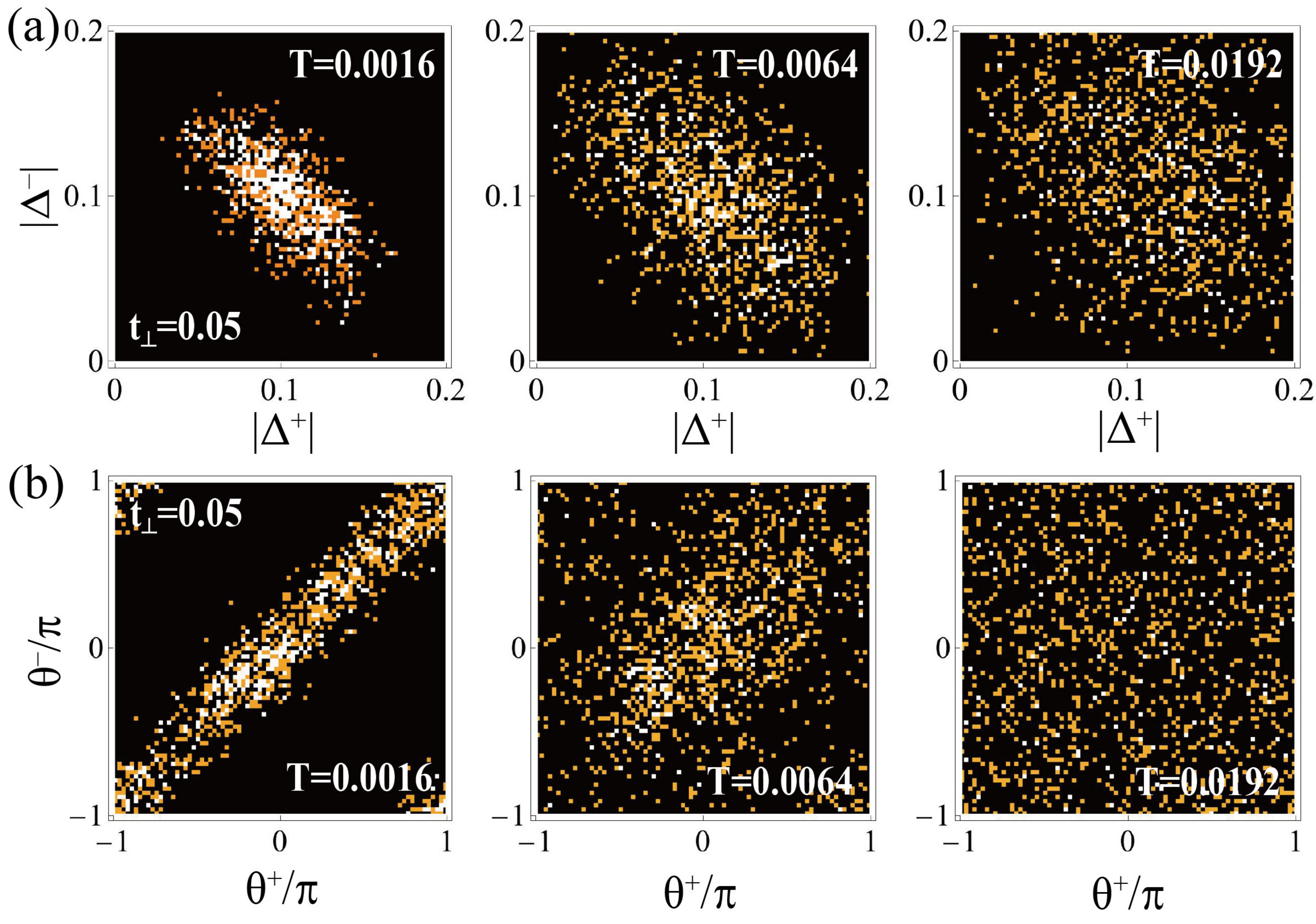}
	\end{center}
	\caption{\textbf{Effect of the interlayer hopping.} Intensity plot of the joint distribution of (A) the pairing amplitudes $|\Delta^{\pm}|$ and (B) their phases $\theta^\pm$ of the trilayer model, showing strong fluctuations even deep inside the superconductivity for a finite $t_\perp=0.05$. Other parameters are $V=0.15$ and $J=0.5$, with $T_c\approx0.008$ reduced slightly from $0.009$ at $t_\perp=0$.}
	\label{fig3}
\end{figure}

A nonzero interlayer hopping introduces a small but finite $h\propto t_\perp^2 \propto \delta_h^2$, which plays the role of a transverse field in the GL free energy and breaks the global SU(2) symmetry. For $h>0$ ($<0$), the GL free energy is minimized at $\theta^+=\theta^-$ ($\theta^+-\theta^-=\pi$) and $|\Delta^+|=|\Delta^-|$, with Gaussian fluctuations of both $|\Delta^+|-|\Delta^-|$ and $\theta^+-\theta^-$ proportional to $\delta_h^{-1}$, as derived in the supplemental information. Thus, the fluctuations remain strong for small $\delta_h$. This is confirmed in Figures \ref{fig3}A and \ref{fig3}B  from our numerical calculations of the joint distribution of $|\Delta^\pm|$ and $\theta^\pm$ for $t_\perp=0.05$ and $V=0.15$. Indeed, while the data get more clustered around the GL solution with lowering temperature, strong fluctuations persist even at $T=0.0016$ well below the BKT transition, indicating the lack of coherence along the $c$-axis. By rewriting $\mathcal{T}^\Delta_x=|\Delta^+\Delta^-|\cos(\theta^+-\theta^-)$, we find an intrinsic Josephson coupling within the trilayer structure, which predicts a Josephson frequency (if exist) only half that for intralayer pairing but is absent in the bilayer model (see supplemental information). We suggest future experiments to examine to what extent these features may be realized in real materials.

\subsection*{Pairing symmetry}

\begin{figure}
	\begin{center}
		\includegraphics[width=14cm]{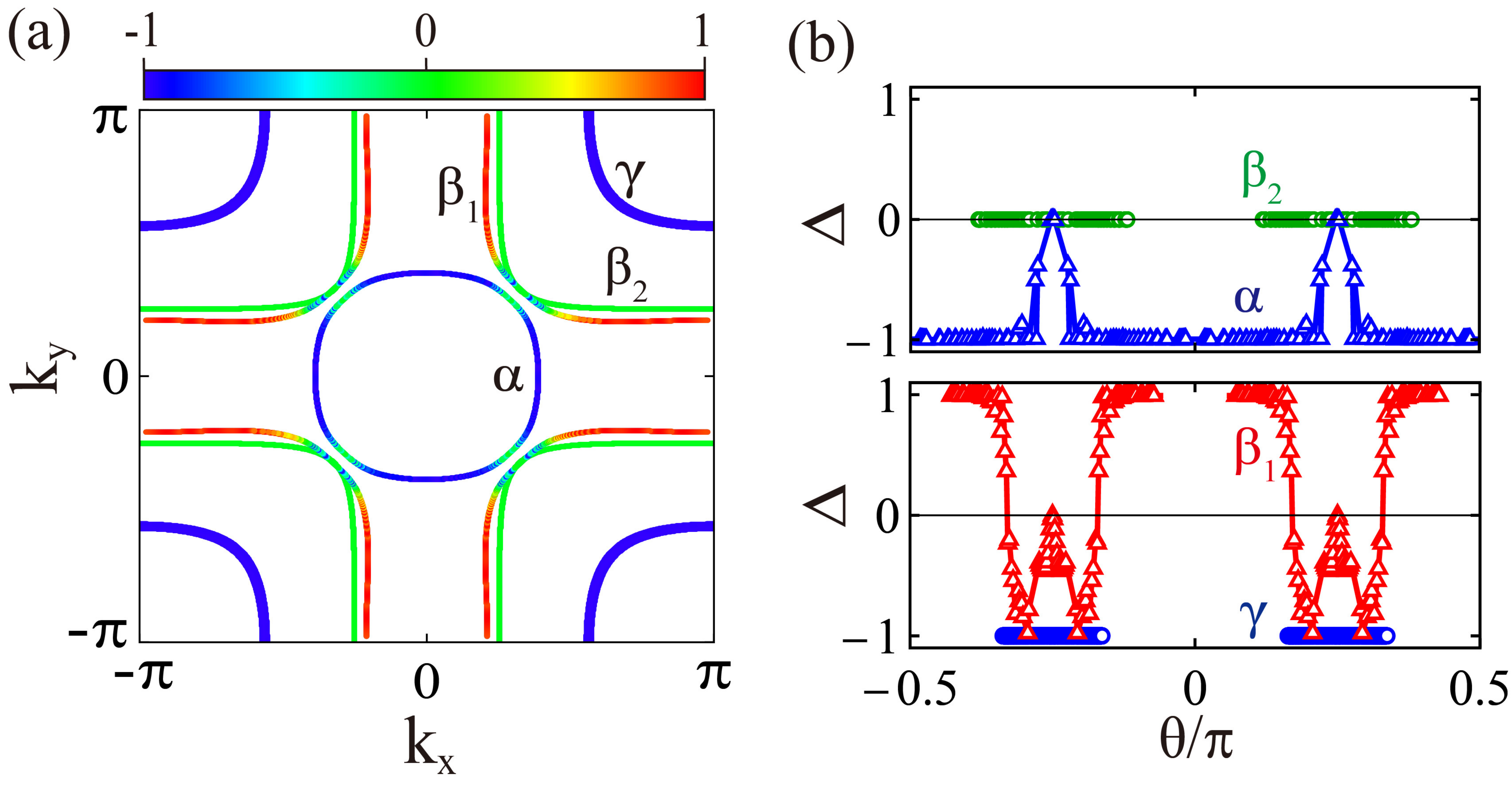}
	\end{center}
	\caption{\textbf{Superconducting gap structures.} (A) Illustration of some typical gap structures (normalized to their maximum values) on four Fermi pockets predicted by first-principles calculations. (B) Their variations with the azimuthal angle $\theta$ within $(-\pi/2, \pi/2)$.}
	\label{fig4}
\end{figure}

The mean-field gap structures in momentum space may be tentatively obtained by constructing a Bardeen-Cooper-Schrieffer (BCS) Hamiltonian with the static pairing fields $\Delta^a$ and realistic tight-binding parameters\upcite{Luo2024}. First-principles calculations have predicted four Fermi pockets denoted as $\alpha$, $\beta_{1/2}$, and $\gamma$ in Figure \ref{fig4}A. Among them, $\gamma$ comes mainly from the $d_{z^2}$ bonding orbital, and all three others are hybridization bands of both $d_{z^2}$ and $d_{x^2-y^2}$ characters. One may roughly think of the two outer layers to first form bonding and antibonding orbitals. Their antibonding orbitals give rise to the $\beta_2$ Fermi surface, while their bonding orbitals further hybridize with the inner layer to form a bonding combination ($\alpha$) and an antibonding combination ($\beta_1$). Such bonding/antibonding characters determine most of their gap properties except for some narrow regions. The BCS Hamiltonian can be diagonalized to project out the gaps on each Fermi pocket. Figures~\ref{fig4}A and \ref{fig4}B plot some typical results and we find extended $s^\pm$-wave gaps with (accidental) nodes along and close to the zone diagonal. Specifically, the gap on $\gamma$ is nodeless, the gap on $\alpha$ has the same sign as $\gamma$ but contains nodes along the zone diagonal where the $d_{x^2-y^2}$-$d_{z^2}$ hybridization is zero, and the gap on $\beta_1$ has an opposite sign on most regions of the Fermi surface with nodes along the zone diagonal but, depending on model details, may also change sign to give some accidental nodes near the zone diagonal. Notably, $\beta_2$ remains metallic because of its antibonding character formed of two outer layers. Some of these details may change if other paring channels are included. For example, introducing additional interlayer $d_{x^2-y^2}$ pairing might eliminate the sign change and accidental nodes on $\beta_1$, while direct interlayer pairing of $d_{z^2}$ between two outer layers would open a superconducting gap on $\beta_2$. But the main features, namely the extended $s^{\pm}$-wave pairing symmetry and the gap nodes (or minima) along the zone diagonal, are robust as long as the $d_{z^2}$ interlayer pairing interaction dominates.

\section*{DISCUSSION}

We have proposed an interlayer pairing scenario for La$_4$Ni$_3$O$_{10}$ and derived a reduced maximum ratio $T_c/J\approx 0.02-0.03$ in the trilayer model from $0.04-0.05$ in the bilayer model. We show that this reduction may be a natural consequence of interlayer pairing due to the competition of two outer layers to form spin-singlet pairs with the inner layer respectively. This induces some kind of frustration and enhances the superconducting fluctuations between layers. Similar simulations for intralayer pairing have been performed for cuprates, but the maximum ratio $T_c/J$ seems robust except for some very special situations\upcite{Qin2024}. It has been argued that the observed enhancement of maximum $T_c$ in multilayer cuprates might actually come from the increase of the intralayer superexchange interaction or tuning to reach the optimal condition for maximizing $T_c/J$. While such details may be debated, our systematic comparison here reveals a probably fundamental distinction between two systems associated with their potentially different (interlayer versus intralayer) pairing mechanisms. An alternative explanation would be demanded if intralayer pairing governs bilayer and trilayer nickelate superconductors.

Lately, it has been reported that La$_3$Ni$_2$O$_7$ may also adopt a single layer-trilayer alternate (1313) structure\upcite{Puphal2023,Chen2023h,Wang2023f}, raising concerns regarding the crystal structure truly responsible for its high-temperature superconductivity. While some recent experiments have already provided supportive evidences for the bilayer structure\upcite{Li2024b}, our calculations here imply that the (1313) structure cannot give a maximum $T_c$ as high as that in the bilayer structure. 

All together, our work reveals potentially rich physics in multilayer superconductors with interlayer pairing, which may be distinctively different from that for intralayer pairing but has not been well noticed before. This points out a promising novel direction for future investigations.


\clearpage

\section*{FUNDING AND ACKNOWLEDGEMENTS}

This work was supported by the Strategic Priority Research Program of the Chinese Academy of Sciences (Grant No. XDB33010100), the National Natural Science Foundation of China (Grants No. 12474136 and No. 12304174), and the National Key Research and Development Program of China (Grant No. 2022YFA1402203). The funders had no role in study design, data collection and analysis, decision to publish or preparation of the manuscript.

\section*{AUTHOR CONTRIBUTIONS}
Y.Y. conceived the idea and directed the research. Q.Q. and J.W. performed the calculations. Q.Q., J.W., and Y.Y. wrote the paper.

\end{document}


\title{Frustrated superconductivity and intrinsic reduction of $T_c$ in trilayer nickelate\\
	\vspace{0.2cm}
- Supplemental Material -}
\author{Qiong Qin}
\affiliation{Beijing National Laboratory for Condensed Matter Physics and Institute of
	Physics, Chinese Academy of Sciences, Beijing 100190, China}
\affiliation{University of Chinese Academy of Sciences, Beijing 100049, China}
\author{Jiangfan Wang}
\affiliation{School of Physics, Hangzhou Normal University, Hangzhou, Zhejiang 311121, China}
\author{Yi-feng Yang}
\email[]{yifeng@iphy.ac.cn}
\affiliation{Beijing National Laboratory for Condensed Matter Physics and Institute of
Physics, Chinese Academy of Sciences, Beijing 100190, China}
\affiliation{University of Chinese Academy of Sciences, Beijing 100049, China}
\affiliation{Songshan Lake Materials Laboratory, Dongguan, Guangdong 523808, China}

\maketitle

\subsection{A. Derivation of the effective action}
We use the coherent-state path integral to transform the Hamiltonian to the following action:
\begin{eqnarray}
	S&=&\int_{0}^{\beta}d\tau\left\{ \sum_{lijs}\bar{c}_{lis}(\tau)\left[(\partial_{\tau}-\mu)\delta_{ij}-t_{ij}\right]c_{ljs}(\tau)+\sum_{lis}\bar{d}_{lis}(\tau)\partial_{\tau}d_{lis}(\tau)\right.\notag\\
	&&\left.-\sum_{lijs}V_{ij}\left(\bar{d}_{lis}(\tau)c_{ljs}(\tau)+c.c. \right)-t_\perp\sum_{ais} \left(\bar{d}_{0is}(\tau) d_{ais}(\tau)+c.c.\right)-\frac{3J}{4}\sum_{ai}\bar{\Phi}_{i}^{a}(\tau)\Phi_{i}^{a}(\tau)\right\}.
\end{eqnarray}
where $\Phi_{i}^a(\tau)=\frac{1}{\sqrt{2}}\left[d_{0i\downarrow}(\tau)d_{ai\uparrow}(\tau)-d_{0i\uparrow}(\tau)d_{ai\downarrow}(\tau)\right]$ denotes the local interlayer spin-singlet pairing of $d_{z^2}$ at site $i$ between the inner layer and the $a$-th outer layer.

The auxiliary interlayer pairing fields $\Delta_i^a(\tau)$ are then introduced to decouple the pairing term:
\begin{equation}
-\frac{3J}{4}\bar{\Phi}_{i}^{a}(\tau)\Phi_{i}^{a}(\tau)\rightarrow \sqrt{2}\bar{\Delta}^a_i(\tau)\Phi_{i}^a(\tau)+\sqrt{2}\bar{\Phi}^a_{i}(\tau)\Delta^a_i(\tau)+\frac{8\bar{\Delta}^a_i(\tau)\Delta^a_i(\tau)}{3J},
\end{equation}
where $\bar{\Delta}^a_i(\tau)$ is the complex conjugate of $\Delta^a_i(\tau)$. In the static auxiliary field approximation, we ignore the  imaginary time dependence of the auxiliary fields, $\Delta^a_i(\tau)\rightarrow\Delta^a_i$, but retain their spatial dependence.

Upon Fourier transformation, the above action becomes
\begin{eqnarray}
S=\sum_{n}\bar{\psi}_n(-i\omega_n+O)\psi_n+\frac{8\beta}{3J}\sum_{ia}|\Delta_i^a|^2,\label{Stp}
\end{eqnarray}
where $\bar{\psi}_n=\left(\bar{c}_{+\uparrow}, c_{0\downarrow}, \bar{c}_{-\uparrow}, c_{+\downarrow}, \bar{c}_{0\uparrow}, c_{-\downarrow}, \bar{d}_{+\uparrow}, d_{0\downarrow}, \bar{d}_{-\uparrow}, d_{+\downarrow}, \bar{d}_{0\uparrow}, d_{-\downarrow}\right)$ in which $\bar{c}_{ls}$ and $\bar{d}_{ls}$ are both row vectors of length $N$ (the lattice size) denoting $(\bar{c}_{l1s} (\tilde{s}i\omega_n),\cdots, \bar{c}_{lNs}(\tilde{s}i\omega_n))$ and $(\bar{d}_{l1s} (\tilde{s}i\omega_n),\cdots, \bar{d}_{lNs}(\tilde{s}i\omega_n))$, respectively. Here $\tilde{s}=1$ for $s=\uparrow$ and $\tilde{s}=-1$ for $s=\downarrow$. The matrix $O$ takes the form:
\begin{eqnarray}
O=\left(\begin{array}{cc}
A_1&B_1\\
B_1&D_1
\end{array}\right),~~A_1=\left(\begin{array}{cc}
A&0\\
0&-A\\
\end{array}\right),~~B_1=\left(\begin{array}{cc}
B&0\\
0&-B\\
\end{array}\right),~~D_1=\left(\begin{array}{cc}
M&D\\
-D&M^*\\
\end{array}\right),
\end{eqnarray}
with
\begin{eqnarray}
A=\left(\begin{array}{ccc}
-T&0&0\\
0&T&0\\
0&0&-T\\
\end{array}\right),~~B=\left(\begin{array}{ccc}
-\mathcal{V}&0&0\\
0&\mathcal{V}&0\\
0&0&-\mathcal{V}\\
\end{array}\right),~~D=\left(\begin{array}{ccc}
0&-T_\perp&0\\
T_\perp&0&T_\perp\\
0&-T_\perp&0\\
\end{array}\right),~~M=\left(\begin{array}{ccc}
0&M_+&0\\
M_+^*&0&M_-^*\\
0&M_-&0\\
\end{array}\right),
\end{eqnarray}
where $T_{ij}=t_{ij}+\mu\delta_{ij}$, $(T_\perp)_{ij}=t_\perp\delta_{ij}$, $\mathcal{V}_{ij}=V_{ij}$, and $(M_a)_{ij}=\Delta_i^a\delta_{ij}$ with $i, j=1,\cdots, N$.

For numerical simulations, we integrate out the fermionic degrees of freedom and obtain the effective action of the auxiliary pairing fields:
\begin{equation}
S_{\rm eff}(\{\Delta^a_i\})=\frac{8\beta}{3J}\sum_{ia}\bar{\Delta}_i^{a}\Delta_i^{a}-\sum_{i}\ln(1+e^{-\beta\tilde{\Lambda}_i})-\sum_{i}\ln(1+e^{-\beta\Lambda_i}),
\end{equation}
where $\tilde{\Lambda}_i$ and $\Lambda_i$ are the eigenvalues of the matrix $A_1$ and $Q=B_1A_1^{-1}B_1+D_1$ to reduce the computational time. This is exactly Eq. (3) in the main text.

When $t_{\perp}=0$, $D_1$ is diagonal and the above procedure can be further simplified to give
\begin{equation}
S=\sum_{n}\bar{\psi}_{1n}(-i\omega_n+O_1)\psi_{1n}+\sum_{n}\bar{\psi}_{2n}(-i\omega_n+O_2)\psi_{2n}+\frac{8\beta}{3J}\sum_{ia}|\Delta_i^a|^2,\label{Stp0}
\end{equation}
where 
\begin{eqnarray}
&&\bar{\psi}_{1n}=\left(\bar{c}_{+\uparrow}(i\omega_n), c_{0\downarrow}(-i\omega_n), \bar{c}_{-\uparrow}(i\omega_n), \bar{d}_{+\uparrow}(i\omega_n), d_{0\downarrow}(-i\omega_n), \bar{d}_{-\uparrow}(i\omega_n)\right),\\
&&\bar{\psi}_{2n}=\left(c_{+\downarrow}(-i\omega_n), \bar{c}_{0\uparrow}(i\omega_n), c_{-\downarrow}(-i\omega_n), d_{+\downarrow}(-i\omega_n), \bar{d}_{0\uparrow}(i\omega_n), d_{-\downarrow}(-i\omega_n)\right).
\end{eqnarray}
Again, $\bar{c}_{ls}(i\omega_n)$ and $\bar{d}_{ls}(i\omega_n)$ are row vectors of length $N$ as defined above. The matrices $O_1$ and $O_2$ are:
\begin{eqnarray}
O_1=\left(\begin{array}{cc}
A&B\\
B&M
\end{array}\right),~~~O_2=\left(\begin{array}{cc}
-A&-B\\
-B&M^*
\end{array}\right).
\end{eqnarray}
Integrating out the fermionic degrees of freedom gives the effective action:
\begin{equation}
\begin{split}
S_{\rm eff}(\{\Delta^a_i\})=&\frac{8\beta}{3J}\sum_{ia}\bar{\Delta}_i^{a}\Delta_i^{a}-\sum_{i;s=\pm}\ln(1+e^{-s\beta\Lambda_i^{0}})-\sum_{i;s=\pm}\ln(1+e^{-\beta\Lambda_i^s}),
\end{split}
\end{equation} 
where $\Lambda_i^0$, $\Lambda_i^+$, and $\Lambda_i^-$ are the eigenvalues of $A$, $Q_1=BA^{-1}B+M$, and $Q_2=-BA^{-1}B+M^*$, respectively.

\subsection{B. Effect of imbalance between inner and outer layers}
\begin{figure}[h]
	\begin{center}
		\includegraphics[width=10.5cm]{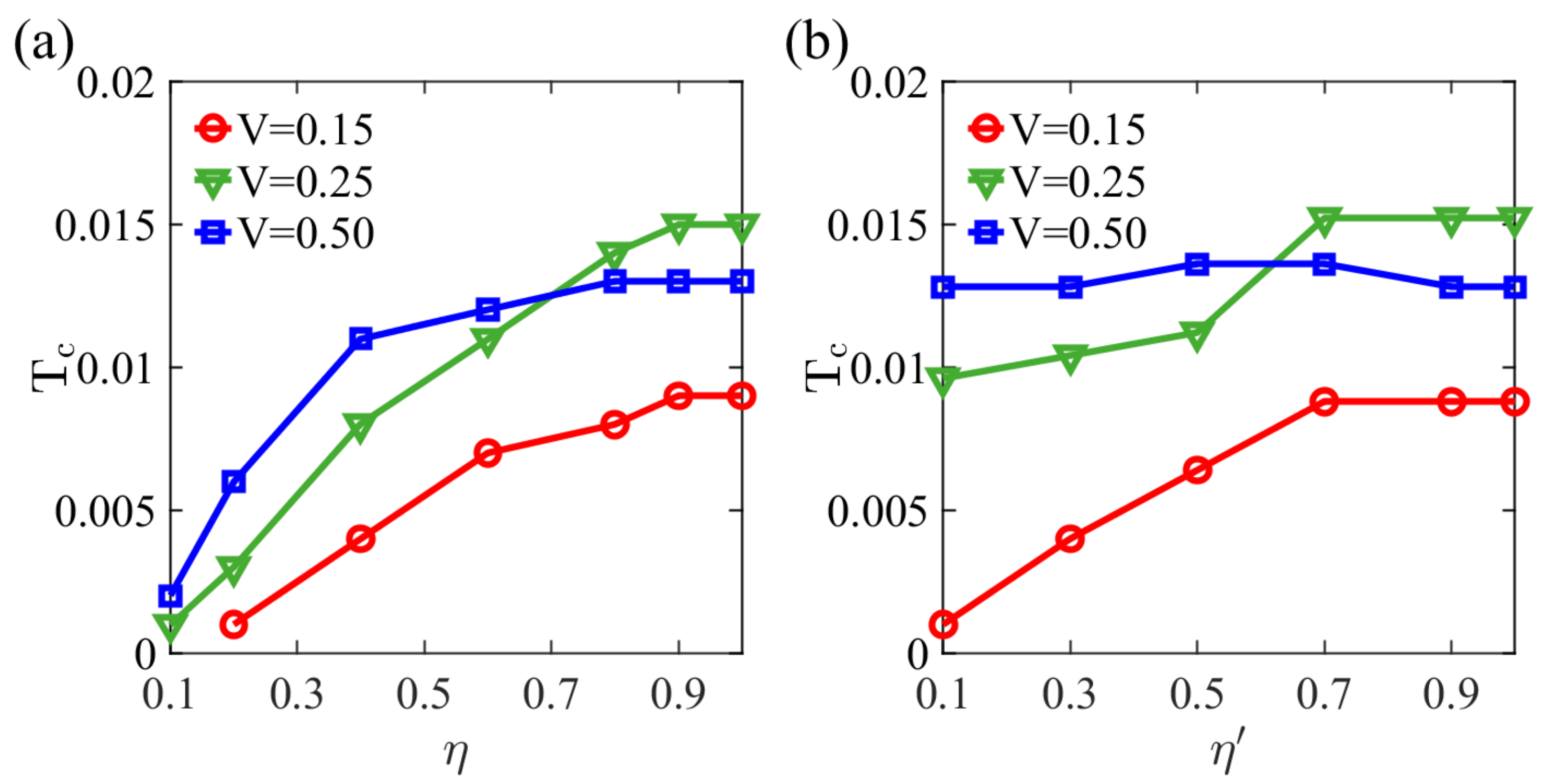}
	\end{center}
	\caption{Evolution of $T_c$ as functions of (a) $\eta=V_{\rm outer}/ V_{\rm inner}$ and (b) $\eta'=\mu_{\rm outer}/ \mu_{\rm inner}$ for three typical values of $V=V_{\rm inner}$ with $\mu_{\rm inner}=-1.3$ and $J=0.5$. }
	\label{figS2}
\end{figure}

We have also performed numerical simulations by assuming different hybridizations or chemical potentials for the inner and two outer layers. Figure \ref{figS2}(a) shows $T_c$ as a function of $\eta=V_{\rm outer}/ V_{\rm inner}$ for three typical values of $V=V_{\rm inner}$. We find  $T_c$ decreases rapidly as $\eta$ decreases from unity and drops to zero as $\eta$ approaches zero. The latter reflects the key role of the outer layers in causing the superconductivity. Figure \ref{figS2}(b) plots $T_c$ as a function of $\eta'=\mu_{\rm outer}/ \mu_{\rm inner}$ for a fixed $\mu_{\rm inner}=-1.3$ and the same values of $V$. We see $T_c$ decreases as $\eta'$ decreases for small $V$ but remains less affected for large $V$.

\subsection{C. Derivation of the Ginzburg-Landau free energy}

The GL free energy density is related to the effective action by $f_\text{GL}=S_{\rm eff}/\beta N$. To simplify the derivation, we treat $t_\perp$ as a perturbation such that
\begin{equation}
f_\text{GL}=f_\text{GL}^{(0)}+ f_\text{GL}^{(2)}+O(t_\perp^4).
\end{equation}
For $t_\perp=0$, Eq. (\ref{Stp0}) gives
\begin{equation}
 f_\text{GL}^{(0)}=\frac{8}{3JN}\sum_{ia}|\Delta_{i}^a|^2-\frac{1}{\beta N}\sum_n\text{Tr}\ln\left(-i\omega_n+O_1\right)-\frac{1}{\beta N}\sum_n\text{Tr}\ln\left(-i\omega_n+O_2\right).
\end{equation}
Since $\Delta_i^a$ only enter $M$ and $M^*$, we may further make the expansion:
\begin{eqnarray}
&&\text{Tr}\ln\left(-i\omega_n+O_1\right)=	\text{Tr}\ln \left(\begin{array}{cc}
	-i\omega_n+A  & B \\
	B & -i\omega_n
\end{array}\right) -\frac{1}{2}\text{Tr}(P_1M)^2-\frac{1}{4}\text{Tr}(P_1M)^4, \notag \\
&&\text{Tr}\ln\left(-i\omega_n+O_2\right)=	\text{Tr}\ln \left(\begin{array}{cc}
	-i\omega_n-A  & -B \\
	-B & -i\omega_n
\end{array}\right)-\frac{1}{2}\text{Tr}(P_2M^*)^2-\frac{1}{4}\text{Tr}(P_2M^*)^4,
\end{eqnarray}
where
\begin{equation}
P_1=\left(\begin{array}{ccc}
		P_+ &0 &0 \\
		0 & P_- &0 \\
		0 & 0 & P_+ 
		\end{array}\right),\quad P_2=\left(\begin{array}{ccc}
		P_- &0 &0 \\
		0 & P_+ &0 \\
		0 & 0 & P_- 
	\end{array}\right), \quad P_\pm=\frac{1}{-i\omega_n+\mathcal{V}(i\omega_n\pm T)^{-1} \mathcal{V}}.
\end{equation}
This gives the free energy density (up to a constant):
\begin{eqnarray}
	f_\text{GL}^{(0)}=&&\frac{8}{3JN}\sum_{ia}|\Delta_{i}^a|^2+\frac{1}{\beta N}\sum_n\text{Tr}\left[(P_-M_+^*P_+M_+)^2+(P_-M_-^*P_+M_-)^2 +2P_-M_+^*P_+M_+ \right. \notag\\
&&+2P_-M_-^*P_+M_- +P_-M_+^*P_+M_+P_-M_-^*P_+M_- \left.+P_-M_+^*P_+M_-P_-M_-^*P_+M_+\right]. \label{eq:FGL}
\end{eqnarray}
Assuming spatially uniform order parameters, $(M_a)_{ii}=\Delta_i^a=\Delta^a$, the above formula can be simplified to
\begin{equation}
	f_\text{GL}^{(0)}=c_1'\left(|\Delta^+|^2+|\Delta^-|^2\right)+c_2\left(|\Delta^+|^2+|\Delta^-|^2\right)^2= c_1'\Psi^\dagger \Psi +c_2 (\Psi^\dagger \Psi)^2,
\end{equation}
where $\Psi\equiv(\Delta^+,\Delta^-)^T$ and 
\begin{equation}
	c_1'=\frac{8}{3J}+\frac{2}{\beta N}\sum_n\text{Tr}(P_+P_-),~~~~~c_2=\frac{1}{\beta N}\sum_n\text{Tr}(P_+P_-P_+P_-).
\end{equation}
For $c_1'<0$ and $c_2>0$, $f_\text{GL}^0$ is minimized at $\Psi^\dagger \Psi=-c_1'/2c_2$.

To obtain $f_\text{GL}^{(2)}$, we expand the effective action (\ref{Stp}) to $O(t_\perp^2)$ and obtain
\begin{equation}
	f_\text{GL}^{(2)}=-\frac{2t_\perp^2}{\beta N}\sum_n\text{Tr}\left[P_-^2 M_+^*P_+^2M_++P_-^2 M_-^*P_+^2M_+P_-^2 M_+^*P_+^2M_-+P_-^2 M_-^*P_+^2M_-\right],
\end{equation} 
which, for uniform order parameters $\Delta^a$, reduces to
\begin{eqnarray}
	f_\text{GL}^{(2)}=-2t_\perp^2 c_3\left[|\Delta^+|^2+|\Delta^-|^2+(\bar{\Delta}^-\Delta^++c.c.)\right]
\end{eqnarray}
with 
\begin{equation}
	c_3=\frac{1}{\beta N}\sum_n\text{Tr}(P_+P_+P_-P_-).
\end{equation}
Putting together, we get the approximate GL free energy density:
\begin{equation}
	f_\text{GL}=c_1\Psi^\dagger \Psi+c_2(\Psi^\dagger \Psi)^2-h\Psi^\dagger \sigma_x \Psi, \label{eq:fGL}
\end{equation}
where $c_1=c_1'-h$ and $h=2t_\perp^2c_3$ plays the role of a transverse magnetic field along the $x$ direction. Introducing the pseudospin $\mathcal{T}^\Delta=\frac{1}{2}\Psi^\dagger \boldsymbol{\sigma}\Psi$ gives Eq. (4) in the main text.

To find the global minimum of Eq. (\ref{eq:fGL}), we write explicitly, $-h\Psi^\dagger \sigma_x\Psi=-2h|\Delta^+||\Delta^-|\cos(\theta^+-\theta^-)$, which is nothing but the Josephson coupling energy of two superconducting layers. Minimizing the free energy with respect to the phase difference $\Delta\theta\equiv \theta^+-\theta^-$ yields the solution $\Delta\theta=0$ for $h>0$ and $\Delta\theta=\pi$ for $h<0$, at which the GL free energy density becomes
\begin{equation}
	f_\text{GL}=c_1\left(|\Delta^+|^2+|\Delta^-|^2\right)+c_2\left(|\Delta^+|^2+|\Delta^-|^2\right)^2- 2|h||\Delta^+||\Delta^-|, \label{eq:fGLh}
\end{equation}
with the solution:
\begin{eqnarray}
	|\Delta^+|=|\Delta^-|=\left\{\begin{array}{cc}
		\Delta_0\equiv\sqrt{\frac{|h|-c_1}{4c_2}}, & ~~~c_1<|h| \\
        0, & ~~~c_1\ge|h|
		\end{array}\right..
\end{eqnarray}
Our numerical simulations of the current trilayer model always find the solution $\Delta\theta=0$ ($h>0$) at low temperatures. Expanding Eq. (\ref{eq:fGL}) around $|\Delta^+|=|\Delta^-|=\Delta_0$ and $\Delta\theta=0$ yields
\begin{equation}
	\delta f_\text{GL}=4c_2\Delta_0^2\left(|\Delta^+|+|\Delta^-|-2\Delta_0\right)^2+h\left(|\Delta^+|-|\Delta^-|\right)^2+h\Delta_0^2\left(\theta^+-\theta^-\right)^2,
\end{equation}	 
indicating strong Gaussian fluctuations of $|\Delta^+|-|\Delta^-|$ and $\theta^+-\theta^-$ proportional to $h^{-1/2}$.

\subsection{D. Derivation of the Josephson effect}
To study the Josephson effect, we introduce a scalar potential $\phi_{a}(\tau)$ for two outer layers relative to the inner layer and take into consideration the imaginary time dependence of the global $\theta^a(\tau)$ inside the BKT phase. For simplicity, we ignore their spatial dependence on each layer. These yield additional dynamics of $\theta^a(\tau)$ with the effective action: 
\begin{eqnarray}
	S_{\theta} & =&\int_{0}^{\beta}d\tau\left[ \sum_{ais}\bar{c}_{ais}\left(i\partial_{\tau}\theta^a+ie\phi_a\right)c_{ais}+\sum_{ais}\bar{d}_{ais}\left(i\partial_{\tau}\theta^a+ie\phi_a\right)d_{ais}-2h\sum_{i}|\Delta^+||\Delta^-|\cos(\theta^+-\theta^-)\right] \notag \\
	& =&\int_{0}^{\beta}d\tau\left[ i\sum_{a}\mathcal{N}_{a}\left(\partial_{\tau}\theta^a+e\phi_a\right)-2U\cos(\theta^{+}-\theta^{-})\right], 
\end{eqnarray}
where $\mathcal{N}_a=\sum_{is}\left(\bar{d}_{ais}d_{ais}+\bar{c}_{ais}c_{ais}\right)$ and $U=h\sum_{i}|\Delta^{+}||\Delta^{-}|$. The above action can be further split into two terms, $S_\theta=S_0+S_J$, with
\begin{eqnarray}
S_0&=&\frac{i}{2}\int_{0}^{\beta}d\tau \mathcal{N} \left(\partial_{\tau}\bar{\theta}+e\bar{\phi}\right),\notag\\
S_J&=&\int_{0}^{\beta}d\tau\left[\frac{ i\Delta \mathcal{N}}{2}\left(\partial_{\tau}\Delta\theta+e\Delta\phi\right)-2U\cos\Delta\theta\right],
\end{eqnarray}
where we have defined $\mathcal{N}=\sum_a \mathcal{N}_a$, $\bar{\theta}=\sum_a \theta^a$, $\bar{\phi}=\sum_a\phi_a$, and $\Delta\mathcal{N}=\mathcal{N}_+-\mathcal{N}_-$, $\Delta\theta=\theta^+-\theta^-$, $\Delta\phi=\phi_+-\phi_-$. Obviously, $S_0$ is not relevant, while $S_J$ describes the dynamics of the phase difference and is responsible for the Josephson effect. Minimizing $S_J$ with respect to $\Delta\mathcal{N}$ and $\Delta\theta$ yields the equations of motion:
\begin{equation}
\partial_\tau\Delta\theta  =-e\Delta\phi, ~~~~~~~	i\partial_\tau\Delta\mathcal{N}  =4U\sin\Delta\theta
\end{equation}
After the Wick's rotation $\tau\rightarrow it$, $\phi\rightarrow-i\phi$, we obtain
\begin{equation}
\partial_t\Delta\theta  =-e\Delta\phi, ~~~~~~~	\partial_t\Delta\mathcal{N}  =4U\sin\Delta\theta
\end{equation}
where $\Delta \phi=\phi_{+}-\phi_{-}$ is the voltage difference between two outer layers. The Josephson current is then
\begin{equation}
I_{-\rightarrow+}=(-e)\frac{d\mathcal{N}_{+}}{dt}=-\frac{e}{2}\frac{d(\Delta \mathcal{N})}{dt}=-2eU\sin\Delta\theta=2eU\sin\left(e\Delta\phi t+\alpha\right),
\end{equation}
where $\alpha$ is a constant offset. We have therefore a characteristic oscillation frequency $\omega_J=e\Delta\phi/\hbar$, half that of the usual ac Josephson effect for intralayer pairing. This can be easily understood, because for interlayer pairing, a phase change $d_{ais} \rightarrow d_{ais}e^{i\theta^a}$ yields $\Delta_{i}^a \rightarrow \Delta_i^a e^{i\theta^a}$, while for intralayer pairing, it yields $\Delta_{i}^a \rightarrow \Delta_i^a e^{2i\theta^a}$.